\documentstyle[multicol,aps,psfig]{revtex} 
 
\begin{document} 
 
\draft 
\preprint{LBNL-46795} 
 
\title{Jet Quenching and Azimuthal Anisotropy of Large $p_T$ Spectra in 
Non-central High-energy Heavy-ion Collisions} 
 
\author{Xin-Nian Wang} 
\address{Nuclear Science Division, Mailstop 70-319, 
        Lawrence Berkeley Laboratory\\ 
University of California, Berkeley, California 94720.} 
\date{August 30, 2000} 
\maketitle 
\begin{abstract} 
Parton energy loss inside a dense medium leads to the suppression of 
large $p_T$ hadrons and can also cause azimuthal anisotropy of hadron 
spectra at large transverse momentum in non-central high-energy heavy-ion 
collisions. Such azimuthal anisotropy is studied qualitatively
in a parton model for heavy-ion collisions at the RHIC energies. 
The coefficient $v_2(p_T)$ of the elliptic anisotropy at large $p_T$ is found
to be very sensitive to parton energy loss. It decreases slowly 
with $p_T$ contrary to its low $p_T$ behavior where $v_2$ increases very rapidly 
with $p_T$. The turning point signals the onset of contributions of
hard processes and the magnitude of parton energy loss.
The centrality dependence of $v_2(p_T)$ is shown to be sensitive 
to both size and density dependence of the parton energy loss and the later
can also be studied via variation of the colliding energy. The anisotropy
coefficient $v_2/\varepsilon$ normalized by the spatial ellipticity $\varepsilon$
is found to decrease significantly toward semi-peripheral collisions, 
differing from the hydrodynamic results for low $p_T$ hadrons.
Constrained by the existing WA98 experimental data at the SPS energy
on parton energy loss, both hadron spectra suppression
and azimuthal anisotropy at high $p_T$ are predicted to vanish for $b>$7-8 fm
in $Au+Au$ collisions at $\sqrt{s}=$130-200 GeV when the hadron rapidity 
density per unit area of the initial overlapped region is less than what
is achieved in the central $Pb+Pb$ collisions at the SPS energy.
\end{abstract} 
 
 
\begin{multicols}{2}

\section{Introduction}

In the study of the dense matter that is created in high-energy heavy-ion
collisions, one crucial issue is the degree of thermalization
through secondary scatterings. If thermalization has been, at least partially,
achieved, there should be collective effects developed through the evolution
of the system. Azimuthal anisotropy in hadron spectra or elliptic flow has 
been proposed as a good signature of collective transverse flow in 
relativistic heavy-ion collisions \cite{ollitrault}. Such an elliptic 
flow has been observed in experiments at the BNL/AGS \cite{e877},
CERN/SPS \cite{na49} and most recently BNL/RHIC by the 
STAR experiment \cite{star}. The strength of the elliptic flow, $v_2$,
experimentally defined as the second coefficient in the Fourier 
decomposition of the particle azimuthal distribution \cite{posk} 
with respect to the reaction plane, was found \cite{posk2} to be 
between the limits of low-density rescattering
\cite{heiselberg} and hydrodynamic 
expansion \cite{ollitrault,heinz,shuryak,hirano},
indicating approach to a higher degree of thermalization with increasing
colliding energies. At RHIC energies, one expects to see $v_2$ becoming
closer to the hydrodynamic limit as the initial energy density increases
and the lifetime of the initial dense matter is getting longer. Because
system evolution will eliminate the geometrical anisotropy which generates
the anisotropy in momentum space, the elliptic flow was 
argued \cite{sorge,heinz} to be sensitive to the 
early dynamics of the system.

In hydrodynamic models, the strength of the differential 
elliptic flow, $v_2(p_T)$, increases almost linearly with 
$p_T$ \cite{heiselberg,heinz} (at $p_T\sim 0$, however,$v_2$ increases 
quadratically with $p_T$) because of collective expansion. At 
large transverse momentum, 
the hydrodynamic model will likely cease to be a valid
mechanism for particle production in high energy nuclear collisions.
Instead, particle production at $p_T> 2$ GeV/$c$ will be 
dominated by hard or semihard processes. The hadron spectra in this
region of phase space typically exhibit a power-law behavior and depend
on how an energetic parton propagates through the dense medium
created in the heavy-ion collisions. The corresponding azimuthal 
anisotropy of hadron spectra at high $p_T$ would also have completely
different behavior from hydrodynamic model results.

Recent theoretical studies of fast partons propagating inside a dense
medium all suggest a large energy loss caused by multiple scattering
and induced gluon radiation \cite{gw1,bdmps,bgz}. Of particular
interest is the quadratic dependence of the total energy loss on the
distance of propagation \cite{bdmps} due to the non-Abelian nature
of gluon radiation in QCD. Such a parton energy loss will 
cause the suppression of large $p_T$ hadron spectra \cite{wgprl,wang98} 
in heavy-ion collisions if the lifetime of the dense medium is long
enough to influence the propagation of fast partons.
The suppression will depend on the average distance that partons 
propagate inside the medium during the lifetime of the dense matter.
Since the transverse distance depends on the azimuthal direction of 
the parton propagation in non-central heavy-ion collisions, one should expect
the hadron suppression caused by parton energy loss to depend on the
azimuthal angle with respect to the reaction plane, thus leading to 
azimuthal anisotropy in high $p_T$ hadron spectra. 

When the momentum transfer $\mu$ for each of 
the few scatterings suffered by an 
energetic parton in a finite system is small so that $\mu/p_T\ll 1$,
the effect of the elastic scattering on azimuthal anisotropy in large
$p_T$ hadron spectra is negligible. Only parton energy loss can cause
sizable azimuthal anisotropy for large $p_T$ hadrons. Therefore,
azimuthal anisotropy and hadron spectra suppression at large $p_T$
should accompany each other in high energy heavy-ion collisions.
For a parton energy loss $dE/dx$ that has a weak dependence on the
parton energy, the hadron spectra suppression and the accompanying
azimuthal anisotropy should all decrease with $p_T$ as we shall show
in this paper. This is in sharp
contrast with the effect of hadronic interactions whose cross 
sections, especially for inelastic processes, increase with energy 
and thus lead to increased hadron suppression and azimuthal 
anisotropy at large $p_T$. Therefore, the $p_T$ dependence of 
hadron spectra suppression and azimuthal anisotropy is a unique
signal of parton scattering and energy loss in the early stage of
heavy-ion collisions.

Azimuthal anisotropy due to jet quenching has been
studied \cite{galoyan,snellings} before with the HIJING Monte Carlo 
model \cite{hijing}. In this paper, we will study the azimuthal anisotropy
in hadron spectra at large $p_T$ using a parton model in which one 
incorporates the parton energy loss via modified parton fragmentation 
functions \cite{wang98,whs}. We will study the sensitivity
of the azimuthal anisotropy, its $p_T$ and centrality dependence to the 
parton energy loss $dE/dx$ and in particular the distance and density 
dependence of the parton energy loss.

We should emphasize here that our study in this paper is only 
qualitative. First of all, the hard sphere geometry we use does not
give an accurate description of the spatial anisotropy $\varepsilon$ 
(or ellipticity). However, the ratio $v_2/\varepsilon$ will reduce
the sensitivity to models of geometry. Secondly,
our theoretical understanding of parton
energy loss is only qualitative so far. There are many uncertainties
in the estimate of $dE/dx$ in QCD. The purpose of our study is to 
demonstrate the effect of jet quenching in hadron spectra and the
azimuthal anisotropy for a given averaged $dE/dx$. Furthermore, 
in our model we will neglect both longitudinal and transverse expansion. 
The longitudinal expansion will change the particle density 
and thus the effective total parton energy loss \cite{baier98} 
in the medium while the transverse expansion will influence the 
spatial ellipticity of the medium as the system evolves with time.
Since the parton energy loss in a QGP is estimated to be much larger than
that in a hadronic gas \cite{bdmps}, we assume that the parton energy loss 
happens mostly in the partonic medium. Then an energetic parton is likely
to travel to the outside of the partonic matter before the transverse
expansion changes the spatial ellipticity of the partonic medium 
significantly. In this case, the expansion will just change the overall
average parton energy loss. For a qualitative study in this
paper, we are only concerned with the average parton energy loss
and we will just consider the dependence of $dE/dx$ on the average
particle density in the initial overlapped region.

\section{Hadron spectra at large $p_T$}

Hadron spectra in $pp$, $pA$ and $AA$ collisions have been systematically
studied in a parton model \cite{wang98}. This model extends the collinear
factorized parton model to include intrinsic transverse momentum and its
broadening due to multiple scattering in nuclear matter. The value of the
intrinsic transverse momentum and its nuclear broadening are adjusted once
and the model can reproduce most of the experimental data in $pp$ and $pA$ 
collisions \cite{wang98}. In $AA$ collisions, we model the effect of parton
energy loss by the modification of parton fragmentation functions \cite{whs}
in which the distribution of leading hadrons from parton fragmentation is
suppressed due to parton energy loss. In non-central collisions, the single
inclusive hadron spectra in this parton model are given by,
\begin{eqnarray}
  & & \frac{d\sigma_{AB}}{dyp_Tdp_Td\phi}=K\sum_{abcd} 
  \int_{b_{min}}^{b_{max}} d^2b d^2r
  t_A(r)t_B(|{\bf b}-{\bf r}|) \nonumber \\
  & & \int dx_a dx_b d^2k_{aT} d^2k_{bT} g_A(k_{aT},Q^2,r)
  g_B(k_{bT},Q^2,|{\bf b}-{\bf r}|) \nonumber \\ 
  & & f_{a/A}(x_a,Q^2,r)f_{b/B}(x_b,Q^2,|{\bf b}-{\bf r}|) 
  \nonumber \\
  & &\frac{D_{h/c}(z_c,Q^2,L(\phi))}{\pi z_c}
  \frac{d\sigma}{d\hat{t}}(ab\rightarrow cd). \label{eq:nch_AA}
\end{eqnarray}
The factor $K=1.5 -2.0$ is used to account for higher order corrections.
The nuclear thickness
function $t_A(b)$ is normalized to $\int d^2b t_A(b)=A$ using the Woods-Saxon
form of nuclear distributions. To take into account of the intrinsic transverse
momentum and its nuclear broadening, $g_A(k_{T},Q^2,r)$ is assumed to have a
Gaussian form and its width is parametrized to fit the $pp$ and $pA$ data. The
impact-parameter dependence of $g_A(k_{T},Q^2,r)$ comes from the nuclear 
broadening of the intrinsic transverse momentum. The parton distributions in
nuclei $f_{a/A}(x_a,Q^2,r)$ are assumed to be factorized into the parton
distributions in a nucleon and the impact-parameter dependent nuclear
modification factor,
\begin{eqnarray}
    f_{a/A}(x,Q^2,r)&=&S_{a/A}(x,r)\left[ \frac{Z}{A}f_{a/p}(x,Q^2)\right.
    \nonumber \\
    &+&\left. (1-\frac{Z}{A}) f_{a/n}(x,Q^2)\right]. \label{eq:shd}
\end{eqnarray}
We will use the parameterization of the nuclear modification in 
HIJING \cite{hijing}. Constrained by the existing $pA$ data, 
nuclear modification of parton distributions and $p_T$ broadening 
produce about 10-30\% increase in the $p_T$ spectra in central
$Au+Au$ collisions relative to $pp$ \cite{wang98}. However, they will
not give any azimuthal anisotropy in the hadron spectra. 

\begin{figure} 
\centerline{\psfig{figure=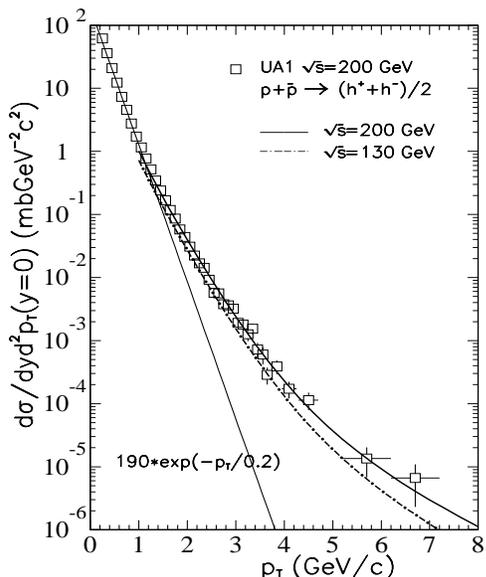,width=2.5in,height=3.0in}}
\caption{Transverse momentum spectra of charged hadrons in $pp$
collisions at $\sqrt{s}=200$ (solid) and 130 GeV (dot-dashed)
as compared to UA1 data \protect\cite{ua1} on $p\bar{p}$ collisions
at $\sqrt{s}=200$ GeV. The thin solid line is a parameterization of the 
data below $p_T<1$ GeV/$c$.}
\label{fig1} 
\end{figure}

Eq.~(\ref{eq:nch_AA}) also applies to  $pp(\bar{p})$ 
collisions in which one simply sets $A=1$ without intrinsic $p_T$ 
broadening and nuclear modification of the parton distributions.
The parton fragmentation functions $D_{h/c}(z_c,Q^2)$ are then given by the 
parameterization of $e^+e^-$ data \cite{bkk}. As an example, we show in 
Fig.~\ref{fig1} the comparison of the parton model calculation (solid line)
with the experimental data \cite{ua1} for $p\bar{p}$ collisions 
at $\sqrt{s}=200$ GeV. We used a $K=1.5$ factor in the calculation. 
The model describes the data very well down to $p_T\sim 1.5$ GeV/$c$. 
Below this scale we expect such a parton model calculation to 
fail and non-perturbative physics to dominate. As an illustration, 
we fit the experimental data below $p_T$=1 GeV/$c$ with an
exponential distribution (thin solid line),
\begin{equation}
  \frac{d\sigma_{soft}}{d\eta d^2p_T}=\frac{\sigma_{in}}{2\pi T_0^2}
  \frac{dn_{ch}}{d\eta} e^{-p_T/T_0}\; , \label{eq:exp}
\end{equation}
with $T_0=0.2$ GeV. The normalization of the fit is
determined by $dn_{ch}/d\eta=2.3$ in the central rapidity region
and $\sigma_{in}=42$ mb at $\sqrt{s}=200$ GeV. One can notice that 
the data differ significantly from the exponential fit already at around 
$p_T=1.5$ GeV/$c$. Beyond this scale, the spectra have a power-law 
behavior characteristic of hard parton scattering. We can therefore use
the parton model to describe the hadron spectra for $p_T>2$ GeV/$c$. 
In Fig.~\ref{fig1}, we also plot the hadron spectra for $pp$ collisions at
$\sqrt{s}=130$ GeV. The spectrum differs very little from that 
at $\sqrt{s}=200$ GeV at intermediate $p_T$ and is only about a factor 2 lower
at $p_T=7$ GeV/$c$. 

To take into account the effect of parton energy loss in $AA$ 
collisions, we use modified effective fragmentation functions, 
$D_{h/c}(z_c,Q^2,L(\phi))$ 
for a produced parton $c$ which has to travel in the medium a distance 
$L(\phi)$ that depends on the azimuthal angle in non-central $AA$ 
collisions. This will be the dominant source of anisotropy in hadron spectra
in this parton model.
We will use a phenomenological model \cite{whs} for the modified 
fragmentation functions. The modification in this model  depends on two
parameters: the energy loss per scattering $\epsilon_c$ and the
mean free path $\lambda_c$ for a propagating parton $c$. The energy
loss per unit length of distance is then $dE_c/dx=\epsilon_c/\lambda_c$. 
We also assume that a gluon's mean free path is half that of a quark
and then the energy loss $dE/dx$ is twice that of a quark. 
According to a pQCD study of the parton energy loss \cite{bdmps},
\begin{equation}
  \frac{dE}{dx}\approx \frac{\alpha_s N_c}{4}\frac{L}{\lambda} \mu^2 \; ,
  \label{eq:dedx}
\end{equation}
where $\alpha_s$ is the strong coupling constant, $N_c=3$ and $\mu$ is
the average transverse momentum kick the propagating parton suffers in
the medium per scattering. So the energy loss per scattering
$\epsilon$ in our model is now related to $\mu$ by
\begin{equation}
  \epsilon=\frac{\alpha_s N_c}{4}L\mu^2 \;.
\end{equation}

\begin{figure} 
\centerline{\psfig{figure=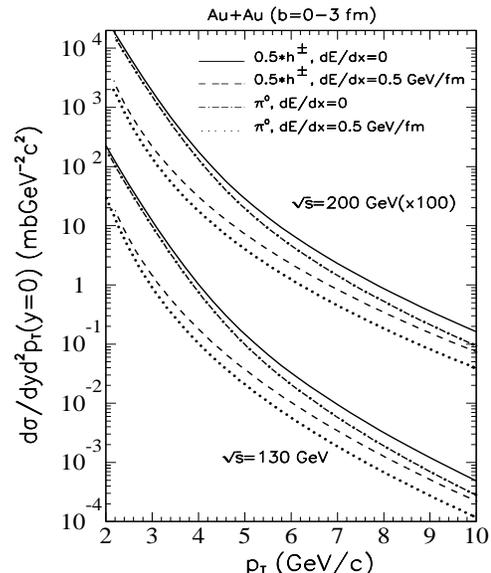,width=2.5in,height=3.0in}} 
\caption{Transverse momentum spectra of charged hadrons (solid) and
$\pi^0$ (dot-dashed) in central $Au+Au$ collisions at $\sqrt{s}=200$ 
and 130 GeV without energy loss and with $dE/dx=0.5$ GeV/fm (dashed 
lines for charged hadrons and dotted lines for $\pi^0$.)}
\label{fig2} 
\end{figure} 

Shown in Fig.~\ref{fig2} are the calculated spectra for charged hadrons,
averaged over azimuthal angle, in
central $Au+Au$ collisions at $\sqrt{s}=130$ and 200 GeV with (dashed lines)
and without (solid lines) parton energy loss. In the spectra we also add
the soft component in Eq.~(\ref{eq:exp}) with $dn_{ch}/d\eta=520$ (700) for
$\sqrt{s}=130$ (200) GeV from HIJING calculations \cite{wgprl00} and
$\sigma_{in}=\pi b_{max}^2$. We used a constant $dE/dx=0.5 $ GeV/fm 
and $\lambda=1$ fm in the calculation to demonstrate the effect of 
parton energy loss. The contribution from the soft component is 
negligible at $p_T>2$ GeV/$c$ as compared to the hard contribution 
when there is no parton energy loss. However, when parton energy loss is 
assumed, the hard contribution is strongly suppressed
so that the soft component
become important at some smaller $p_T$. This is why the suppression of the 
large $p_T$ spectra due to parton energy loss as compared to the spectra
without energy loss becomes smaller at smaller $p_T$. However, with
$dE/dx=0.5$ GeV/fm, the contribution from the soft component again is
negligible at $p_T>3$ GeV/$c$. The suppression is proportional to parton
energy loss. The detailed dependence of the suppression on the form
and values of parton energy loss can be found in Ref.~\cite{wang98}.
Also shown in Fig.~\ref{fig2} are the $\pi^0$ spectra with (dotted lines)
and without (dot-dashed lines) parton energy loss. Notice that $\pi^0$
spectra at high $p_T$ are about 40-30\% lower than $(h^-+h^+)/2$. 
This means that
$\pi^\pm$ only contribute to about 60-70\% of the total charged hadron spectra.
The rest comes from kaons and protons. Kaons and protons spectra have different
sensitivity  to parton energy loss. This is why the suppression factor for
large $p_T$ $\pi^0$ spectra is larger than the charged hadron spectra.
Detailed study of the flavor dependence of the parton energy loss effects
can also be found in Ref.~\cite{wang98}.

\section{Azimuthal anisotropy}

For non-central $A+A$ collisions, the averaged distance a parton travels
through the medium varies with the azimuthal angle and so does the averaged
total parton energy loss. This will give azimuthal
anisotropy in the hadron spectra at large transverse momentum
with respect to the reaction plane. The reaction plane is defined by the 
beam direction and the impact parameter ${\bf b}$ of heavy-ion collisions.
Experimentally, the reaction plane should be determined by low $p_T$ 
particles that constitute the bulk of matter produced in heavy-ion collisions.
Large momentum hadrons will not affect the determination of this reaction
plane.

\begin{figure} 
\centerline{\psfig{figure=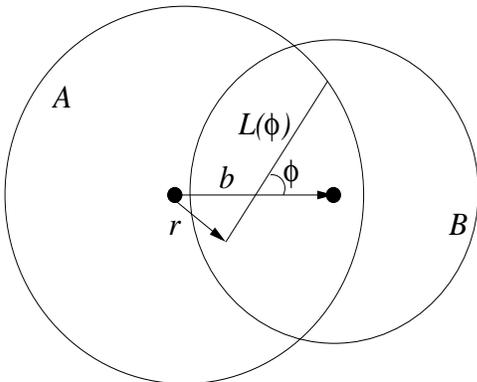,width=2.5in,height=2.0in}} 
\caption{Illustration of the geometry of the overlapped region of
two colliding nuclei $A$ and $B$ in the transverse plane.}
\label{fig:overlap} 
\end{figure}

\begin{figure} 
\centerline{\psfig{figure=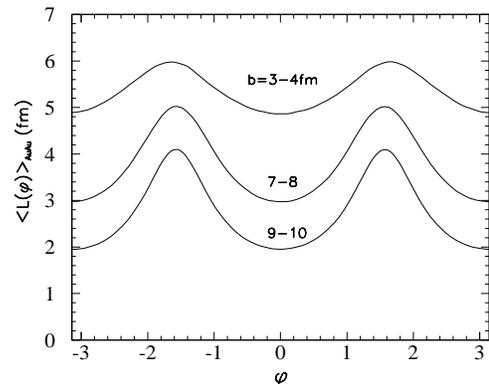,width=2.5in,height=2in}} 
\caption{The averaged distance a fast parton has to travel across 
the overlapped region of two colliding nuclei as a function of the
azimuthal angle $\phi$ for different impact parameters.}
\label{fig:dist} 
\end{figure}

In this model calculation, we will use a hard-sphere nuclear distribution to 
describe the initial geometry of the medium and to estimate the distance 
of parton propagation inside the dense medium. As illustrated by
Fig.~\ref{fig:overlap}, at a given impact parameter $b$,
a parton produced at point ${\bf r}$ has to travel a distance 
$L(\phi,{\bf r},{\bf b})$ inside the overlapped region in the 
azimuthal direction $\phi$. With $L(\phi,{\bf r},{\bf b})$, one can 
then calculate the modified fragmentation function 
$D_{h/c}(z,Q^2,L(\phi,{\bf r},{\bf b}))$ and the hadron spectra.
The cross section is weighted with the overlap function of two 
colliding nuclei according to Eq.~(\ref{eq:nch_AA}).
To help us to understand the azimuthal distribution of hadron spectra,
we show in Fig.~\ref{fig:dist} the averaged distance,
\begin{equation}
\langle L(\phi)\rangle =\frac{1}{T_{AB}(b)}
\int d^2r t_A(r) t_B(|{\rm b}-{\bf r}|) L(\phi,{\bf r},{\bf b}),
\end{equation}
as a function of $\phi$, that the parton has to travel across the 
overlapped region for different impact
parameters, where $T_{AB}(b)=\int d^2r t_A(r) t_B(|{\rm b}-{\bf r}|)$.
As we can see, $\langle L(\phi)\rangle$ on the average decreases with
impact parameter. However, the anisotropy increases toward large impact
parameters. Since jet quenching is directly proportional to 
$\langle L(\phi)\rangle$, one should expect similar azimuthal 
distribution of the hadron suppression at large $p_T$. The azimuthal
anisotropy of the hadron spectra, however, should depend on both the
averaged value of $\langle L(\phi)\rangle$ and its anisotropy.

The azimuthal angle dependence or azimuthal anisotropy 
of the distance $\langle L(\phi)\rangle$ is directly related 
to the spatial deformation of the dense medium which can be characterized
by the spatial ellipticity $\varepsilon$. For a hard-sphere distribution, 
\begin{equation}
\varepsilon(b)\equiv \frac{\langle y^2\rangle - \langle x^2\rangle}
{\langle y^2\rangle + \langle x^2\rangle} = \frac{b}{2R_A}
\end{equation}
is assumed to be given by the initial overlap of two colliding nuclei 
with a radius $R_A\approx 1.12 A^{1/3}$ at impact parameter $b$.
Such a simple geometry is far from realistic. But it is sufficient to
illustrate the qualitative feature of azimuthal anisotropy caused by 
parton energy loss.

\begin{figure} 
\centerline{\psfig{figure=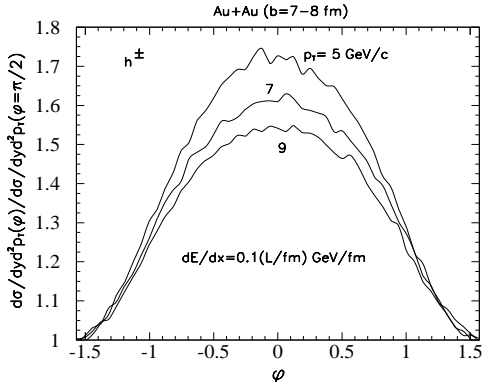,width=2.5in,height=2in}} 
\caption{Azimuthal angle distribution of charged hadrons with
$p_T$=5,7,and 9 GeV/$c$ in semi-peripheral ($b=$ 7-8 fm) $Au+Au$
collisions at $\sqrt{s}=130$ GeV with parton energy loss
$dE/dx=0.1 (L/{\rm fm})$ GeV/fm. Each distribution is normalized
to the spectrum in the direction perpendicular to the reaction plane.}
\label{fig3} 
\end{figure}

Shown in Fig.~\ref{fig3} are the azimuthal angle distributions of
charged hadron spectra at different $p_T$ normalized to the
spectra perpendicular to the reaction plane ($\phi=\pi/2$).
We used $dE/dx=0.1 (L/{\rm fm})$ GeV/fm and $\lambda=1$ fm 
in the calculation. Due to the azimuthal-angle 
dependence of the averaged total energy loss and the consequent
suppression of large $p_T$ hadrons, the hadron spectra show a strong
dependence on the azimuthal angle. The degree of the azimuthal anisotropy
is directly proportional to the hadron suppression at high $p_T$ due to 
parton energy loss. For a fixed value of average total parton energy loss
the relative effect of parton energy loss on the hadron spectra becomes 
smaller at higher $p_T$ \cite{wang98}. That is why the azimuthal 
anisotropy decreases with $p_T$ as shown in Fig.~\ref{fig3}.

\begin{figure} 
\centerline{\psfig{figure=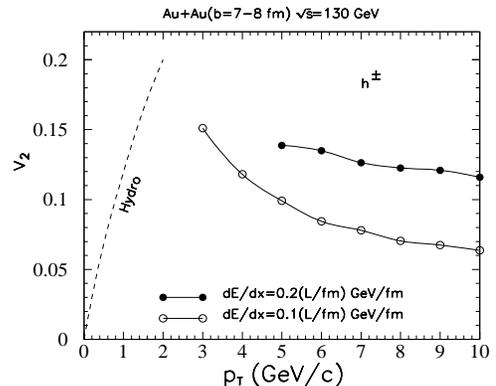,width=2.5in,height=2.0in}} 
\caption{The $p_T$ dependence of the coefficient of 
azimuthal anisotropy $v_2$ for charged hadrons for different values 
of parton energy loss  in semi-peripheral $Au+Au$ collisions at $\sqrt{s}=130$ GeV. 
The dashed line is the hydro result from \protect\cite{heinz}.}
\label{fig4} 
\end{figure}

Following a standard procedure in the study of elliptic flow \cite{posk}, 
we make a Fourier expansion of the hadron distribution in the azimuthal
angle. The elliptic anisotropy coefficient, $v_2$, is defined as the second
order Fourier coefficient,
\begin{equation}
  v_2=\frac{\int_{-\pi}^{\pi} d\phi \cos(2\phi) d\sigma/dyd^2p_Td\phi}
  {\int_{-\pi}^{\pi} d\phi d\sigma/dyd^2p_Td\phi}\; .
\end{equation}
Shown in Fig.~\ref{fig4} are the calculated $v_2$ as functions of the
transverse momentum $p_T$ for different values of the parton 
energy loss $dE/dx$. The elliptic anisotropy coefficient
$v_2$ generally decreases slowly with $p_T$ but increases with $dE/dx$. 
This is in sharp contrast to the hydrodynamic predictions for low
$p_T$ hadrons as shown by the dashed line from Ref.~\cite{heinz}.
We stopped the calculation at $p_T\sim 3-4$ GeV/fm for the given $dE/dx$.
Below these $p_T$ values, the average total parton energy loss 
$\Delta E=\langle L\rangle dE/dx$ will exceed the initial parton
transverse momentum. The hadron spectra from hard processes are
strongly suppressed so that contributions from non-perturbative
processes become dominant and thermal equilibration and hydrodynamics
will determine the elliptic anisotropy. Since the coefficient of the
differential elliptic flow $v_2(p_T)$ for low $p_T$ hadrons increases
with $p_T$, the turning point where $v_2(p_T)$ starts to saturate and
decrease with $p_T$ will provide information on the interplay between hard
and soft components of particle production. Combined with the
magnitude of the azimuthal anisotropy at higher $p_T$, it also
allows one to extract the average parton energy loss in the medium.
Since one cannot make quantitative estimate of parton energy loss, 
we can only give qualitative predictions of $v_2$ and its dependence 
on $p_T$.

\section{Centrality dependence}

\begin{figure}
\centerline{\psfig{figure=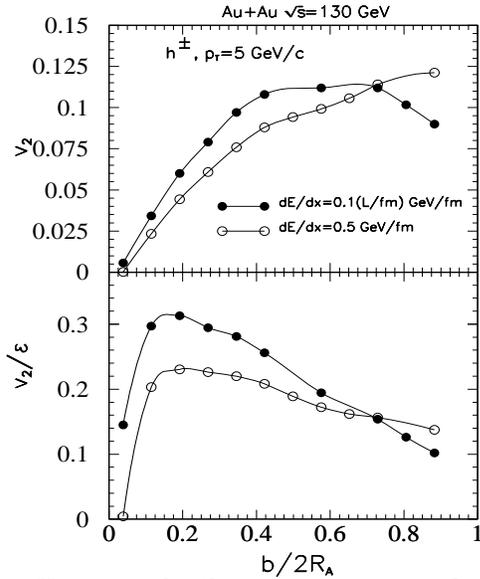,width=2.5in,height=3.0in}} 
\caption{Upper panel: The impact parameter dependence of azimuthal
anisotropy coefficient $v_2$ for charged hadrons with $p_T$=5 GeV/$c$ 
in $Au+Au$ collisions at $\sqrt{s}=130$ GeV for a 
constant $dE/dx=0.5$ GeV/fm (open circle) and distance-dependent
$dE/dx=0.1(L/{\rm fm})$ GeV/fm (solid circle) parton energy loss.
Lower panel: The impact parameter dependence of $v_2/\varepsilon$.
Here, $\varepsilon=b/2R_A$ is the spatial ellipticity 
of the initial overlapped region
of collisions with a hard sphere nuclear distribution.}
\label{fig5} 
\end{figure}

In non-central heavy-ion collisions, the average distance of parton
propagation depends on the impact parameter of collisions. 
The suppression of hadron spectra at large $p_T$ and the elliptic 
anisotropy will then also depend on the centrality. Shown in 
the upper panel of Fig.~\ref{fig5} is the impact parameter dependence of
the calculated $v_2$ at fixed $p_T$=5 GeV/$c$ for two different forms of
parton energy loss, one constant $dE/dx=0.5$ GeV/fm and another
$dE/dx=0.1 (L/{\rm fm})$ GeV/fm with a linear distance dependence. 
In both cases, $v_2$ increases with the impact parameter,
reflecting the increased spatial ellipticity at large impact parameters.
In central collisions, the average 
parton energy loss $dE/dx$ in the second case is larger than the constant 
$dE/dx=0.5$ GeV/fm, leading to a larger $v_2$. However, as the impact
parameter increases further, the average parton energy loss per unit length
with a linear distance dependence
becomes smaller than the constant $dE/dx=0.5$ GeV/fm. This
leads to smaller $v_2$. Eventually, in more peripheral collisions,
$v_2$ decreases with the impact parameter in the second case. Therefore, 
as compared to the case of a constant $dE/dx$, a linear distance 
dependence of $dE/dx$ gives a slower impact parameter dependence of $v_2$.

Since the elliptic anisotropy in hadron spectra is directly related to
the spatial ellipticity, it is useful to study the impact parameter 
dependence of the ratio $v_2/\varepsilon$ as shown in the lower panel of
Fig.~\ref{fig5}. Such a ratio will also reduce the sensitivity to the
modeling of the initial density distribution of the overlapped region,
{\it e.g.}, hard sphere vs. Woods-Saxon nuclear distributions. For both forms 
of parton energy loss, the normalized elliptic anisotropy coefficient 
increases with impact parameter very rapidly in central collisions and then
starts to decrease at relatively small impact parameters. The decrease is
much faster for the case of linear distance dependence of $dE/dx$. This
feature is markedly different from the hydrodynamic models \cite{heinz},
in which $v_2/\varepsilon$ remains constant for a very large range of impact
parameters. So far experimental measurements of the impact parameter
dependence of $v_2$ \cite{e877,na49,star} for low $p_T$ hadrons show a 
behavior roughly similar to the hydrodynamic results. A different
behavior of $v_2/\varepsilon$ at large $p_T$ that decreases with the impact 
parameter will be a clear indication of the effect of parton energy loss.

For a qualitative study in this paper, 
we will neglect the effect of expansion and we are only concerned with 
the average parton energy loss. In this case, the parton energy loss $dE/dx$ 
should depend on the averaged particle density in the overlapped region of 
heavy-ion collisions,
which in turn depends on the centrality of the collisions.

\begin{figure}
\centerline{\psfig{figure=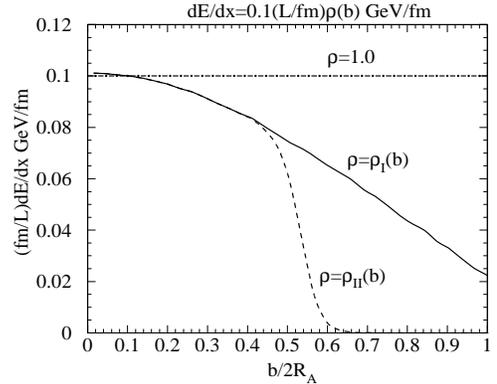,width=2.5in,height=2in}} 
\caption{Three scenarios of impact parameter dependence of parton 
energy loss $dE/dx=0.1 (L/{\rm fm})\rho(b)$ GeV/fm}
\label{fig6} 
\end{figure}

According to Eq.~(\ref{eq:dedx}), $dE/dx$ is proportional to
$\mu^2/\lambda=\mu^2\sigma\rho$ with $\mu$ being the momentum transfer
per scattering. For a parton scattering cross section $\sigma$ that is
screened by $\mu^2$, $\mu^2\sigma$ is roughly a constant. Then $dE/dx$
will be proportional to the parton density $\rho$.
We assume that the initial parton density is proportional to the final
hadron multiplicity per unit transverse area, 
$\rho\sim dN/dy/(\tau_0 S(b))$, with $S(b)$ the transverse 
area of the initial overlapped region and $\tau_0$ the initial time. 
Using the two-component (soft and hard)
model of particle production in HIJING \cite{hijing}, $dN/dy$ has one
term (soft) proportional to the number of participant nucleons $N_{\rm part}$
and another (hard) proportional to the number of binary
collisions $N_{\rm binary}$. In a Glauber model of nuclear collisions,
$S(b)\sim N_{\rm part}^{2/3}(b)$ and $N_{\rm binary}(b)\sim N_{\rm part}^{4/3}(b)$.
Parameterizing the centrality dependence of $dN/dy$ in the HIJING
calculation \cite{wgprl00} for $Au+Au$ collisions at $\sqrt{s}=130$ GeV,
we have
\begin{eqnarray}
  \frac{dE}{dx}&=&\left(\frac{dE}{dx}\right)_0 \rho_I(b); \nonumber \\
  \rho_I(b)&=&0.072  N_{\rm part}^{1/3}(b)
  \left [1+0.13 N_{\rm part}^{1/3}(b)\right ], \label{eq:dedxb}
\end{eqnarray}
where the impact parameter dependence of $N_{part}(b)$ can be calculated
in a Glauber model of nuclear collisions with Woods-Saxon nuclear
distributions. The above centrality dependence of the average
$dE/dx$ is normalized
to central collisions, $dE/dx(b=0)=(dE/dx)_0$, or $\rho(b=0)=1$. 
Such an impact parameter dependence of $dE/dx$ is plotted in
Fig.~\ref{fig6} as the solid line for $(dE/dx)_0=0.1 (L/{\rm fm})$ GeV/fm.

Theoretical studies have predicted a large parton energy loss in a hot
partonic medium as compared to cold nuclear matter \cite{bdmps}.
However, analysis of the large $p_T$ $\pi^0$ spectra in central $Pb+Pb$ 
collisions from WA98 experiment \cite{wa98} at the SPS energy
$\sqrt{s}=17$ GeV shows no indication of parton energy loss \cite{wangprl}.
Since the high $p_T$ spectra at SPS are very sensitive to any change of
the jet cross section, the constraint on parton energy loss by the WA98
data is very stringent and gives a limit that is much smaller
than the most conservative estimate of parton energy loss in a 
hadronic matter \cite{wang98}. This casts 
serious doubts on the accuracy of the theoretical estimates
which are based on a scenario of a static and infinitely large 
($L/\lambda\gg 1$) dense 
partonic gas. Though recent studies \cite{glv,gw,wied} have considered
the finite number of scatterings in a medium with finite size,
it is still not clear whether the estimates are numerically 
consistent with the WA98 data. The experimental
data of WA98 could also imply that the lifetime of the dense partonic system
is too short to induce any parton energy loss.

\begin{figure}
\centerline{\psfig{figure=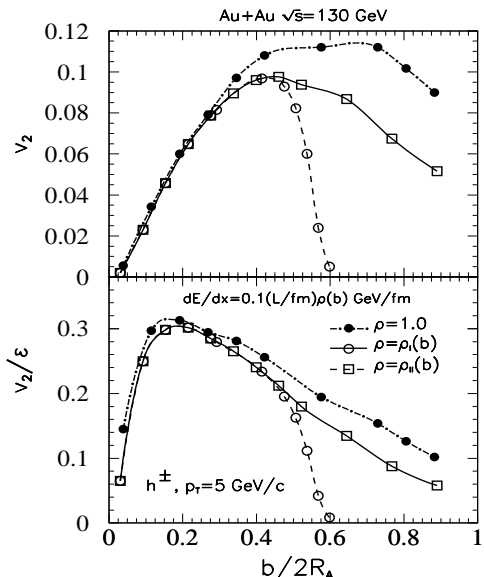,width=2.5in,height=3.0in}} 
\caption{Impact parameter dependence of $v_2$ and $v_2/\varepsilon$ for charged
hadrons with $p_T$=5 GeV/$c$ in $Au+Au$ collisions at $\sqrt{s}=130$ GeV
for three different scenarios of density dependence of parton energy loss.}
\label{fig7} 
\end{figure}

At RHIC energies the initial energy density is expected to be higher than
at the SPS with longer lifetime. Indeed the first experimental 
measurement of $dN_{\rm ch}/d\eta$ by PHOBOS \cite{phobos} 
shows that the particle production for central $Au+Au$
collision at $\sqrt{s}=130 $ GeV is about 50\% higher than in central 
$Pb+Pb$ collisions at the SPS energy, consistent with the 
default value of HIJING simulations. 
If we take the implication of the WA98 data at its face value, we
can assume that the effective parton energy loss vanishes in non-central 
collisions in which the average particle density is equal to or less than
that of central $Pb+Pb$ collisions at the SPS energy. Using HIJING simulations
we find that the particle rapidity density per unit transverse area
in $|\eta|<1$ for $Au+Au$ collisions
becomes less than that of central $Pb+Pb$ collisions at the SPS energy
for impact parameters larger than 7 fm at $\sqrt{s}=130$ GeV, or 8 fm at
$\sqrt{s}=200$ GeV. For smaller impact parameters, we will assume the normal
centrality dependence of the parton energy loss in Eq.~(\ref{eq:dedxb}).
Such a scenario is denoted as density dependence $\rho_{II}(b)$ and
is shown in Fig.~\ref{fig6} as the dashed line. The sudden drop of $dE/dx$ 
with impact parameter might be extreme and is only to illustrate the 
effect of a possible critical behavior in jet quenching.

Shown in Fig.~\ref{fig7} are the calculated coefficients of the azimuthal 
anisotropy $v_2$ and $v_2/\varepsilon$ for charged hadrons at $p_T$=5 GeV/$c$ 
with the above two scenarios of density dependence of parton energy loss
(solid and dashed lines). As a comparison, we also show the results 
of a density-independent $dE/dx$ (dot-dashed lines).
Comparing the results of scenario I of density dependence (solid lines)
to a density-independent $dE/dx$ (dot-dashed lines), it is obvious 
that the density dependence of the parton energy loss $dE/dx$ 
reduces the azimuthal anisotropy at large impact parameters
because of the reduced parton energy loss in peripheral collisions 
as compared to the central ones. The ratio 
$v_2/\varepsilon$ therefore decreases faster toward the peripheral collisions 
as compared to the case of a density-independent $dE/dx$. Because of the 
complication of such a density dependence, it is difficult to single
out the quadratic distance dependence of the parton energy loss by 
studying the centrality dependence of the azimuthal anisotropy alone.
One has to combine it with the study of density dependence by 
measuring the anisotropy for a fixed system at different colliding energies.
In the second scenario of the density dependence of parton energy loss,
$v_2$ will vanish above impact parameter $b>7$ fm, where $dE/dx$ is assumed 
to vanish when the particle density is below what was reached in the central
$Pb+Pb$ collisions at the SPS energy.  The drop should occur at
larger impact parameters for higher colliding energies because of the increased
particle density in the collision region. The behavior of $v_2$ at small or
intermediate impact parameters is dictated by the geometry of the medium
and is not very sensitive to small changes of the density dependence of the
parton energy loss. It is more sensitive to the density dependence at
large impact parameters in more peripheral collisions. The vanishing $dE/dx$
at large impact parameter in the second scenario has very 
dramatic effect on $v_2$. Such a scenario will be easy to verify 
from experimental data. 

Since the azimuthal anisotropy in large $p_T$ hadron spectra in a parton
model is caused predominantly by jet quenching or parton energy loss, such
a study should be complemented with direct measurements of the suppression
of large $p_T$ hadron spectra relative to $pp$ or $pA$ collisions at the same
energy. To facilitate such direct measurements, one needs to scale the
spectra in $AA$ collisions by that of $pp$ and geometric factors. According
to the parton model in Eq.~(\ref{eq:nch_AA}), the ratio
\begin{equation}
  R_{AA}(\langle b\rangle,p_T) \equiv \frac{dN_{AA}/dyd^2p_T}
{\langle N_{\rm binary}\rangle(\langle b\rangle)
dN_{pp}/dyd^2p_T}
\end{equation}
will be 1, independent of $p_T$ and impact parameter $b$ 
if there are no nuclear effects in the large $p_T$ hadron production.
Jet quenching will suppress the large $p_T$ spectra and reduce 
the above ratio for non-vanishing parton energy loss \cite{wang98}. 
Here $\langle N_{\rm binary}\rangle(\langle b\rangle)$, defined as
\begin{equation}
  \langle N_{\rm binary}\rangle(\langle b\rangle)
=\frac{\sigma^{pp}_{in}}{\sigma^{AA}_{in}(b_{min},b_{max})}
\int_{b_{min}}^{b_{max}} d^2b T_{AA}(b),
\end{equation}
is the averaged number of binary collisions that can be calculated 
in a Glauber model of nuclear collisions \cite{wang98} with given
$\langle b\rangle=(b_{min}+b_{max})/2$.

\begin{figure}
\centerline{\psfig{figure=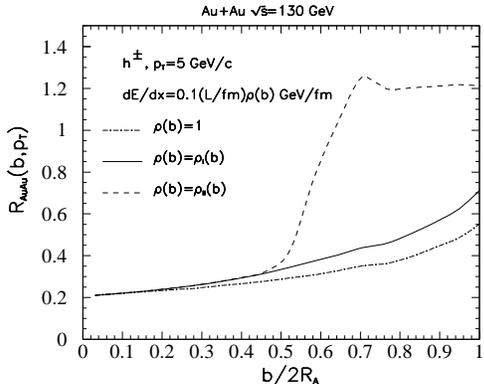,width=2.5in,height=2in}} 
\caption{Impact parameter dependence of the ratio $R_{AA}(p_T,b)$ of the
charged hadron spectra at  $p_T$=5 GeV/$c$ in $Au+Au$ and $pp$ collisions
at $\sqrt{s}=130$ GeV for three different scenarios of density 
dependence of parton energy loss.}
\label{fig8} 
\end{figure}

Shown in Fig.~\ref{fig8} are the
calculated ratios for charged hadron spectra at $p_T=5$ GeV/$c$ in
$Au+Au$ collisions at $\sqrt{s}=130$ GeV as functions of the impact 
parameter with three different density dependencies of parton energy loss.
The suppression due to parton energy loss in general will decrease 
with impact parameter since the size of the overlapped region decreases 
toward more peripheral collisions. For very peripheral collisions, one should
recover the results of $pp$ collisions and the ratio should be 1.
However, if the parton energy loss $dE/dx$ depends on the average particle
density, the ratio (solid) should approach 1 more rapidly than in
the case of a density-independent $dE/dx$ (dot-dashed). Again, if we
assume that parton energy loss vanishes at $b>7$ fm from constraints by
WA98 data at the SPS energy, the ratio will become or is larger 
than 1 already in 
semi-peripheral collisions. The ratio (dashed) will actually become
larger than 1 because of initial $p_T$ broadening or Cronin effect.
Such a centrality dependence of hadron suppression at large $p_T$
combined with the energy dependence will shed light on parton
energy loss in the medium.

\section{Conclusions and Discussions}

In this paper we have studied in a parton model the azimuthal anisotropy 
of large $p_T$ hadron spectra in non-central high-energy heavy-ion collisions
due to parton energy loss. We demonstrated that the anisotropy is
very sensitive to the parton energy loss $dE/dx$ and its dependence
on the size and density of the medium. We predict that the coefficient 
of the anisotropy $v_2$ decreases slowly with $p_T$. The parton model 
also predicts
an early saturation of $v_2$ as a function of the impact parameter $b$.
Once divided by the spatial anisotropy $\varepsilon(b)$, the ratio $v_2/\varepsilon$
will increase with $b$ for very central collisions and then will
decrease toward peripheral collisions. This is in sharp contrast to
the centrality dependence of the elliptic flow $v_2$ for low $p_T$ hadrons.
Hydrodynamic models \cite{heinz} predict almost a constant $v_2/\varepsilon$ 
in a very large range of centralities. We found that the centrality dependence
of $v_2$ is sensitive to the size and density dependence of parton
energy loss $dE/dx$ in the medium. 

Since the azimuthal anisotropy is directly related to parton energy loss, 
one should combine the study of azimuthal anisotropy with the direct 
measurements of hadron spectra suppression at large $p_T$. We propose 
to study the density dependence of parton energy loss by measuring the hadron 
suppression and the anisotropy at different colliding
energies for the same system. 
The centrality dependence of the hadron suppression and anisotropy can then 
provide information on the size dependence of the parton energy loss. 
Perturbative QCD studies predict a nonlinear dependence of the total 
energy loss due to non-Abelian gluon radiation.

Based on constraints by WA98 experimental data on high $p_T$ pion spectra,
we also proposed a density dependence of the parton energy loss that will
vanish in peripheral collisions, where the particle density is equal to or
smaller than what has been achieved in the central $Pb+Pb$ collisions at the
SPS energy. Then both hadron spectra suppression and the azimuthal anisotropy
will disappear in these peripheral collisions. Such a strong
centrality dependence would indicate an onset of parton energy
loss in a dense medium.

Parton energy loss is always associated with momentum
broadening perpendicular to the parton propagation direction. Such 
transverse momentum broadening can also lead to azimuthal anisotropy in
the final hadron spectra. One can characterize the broadening by the
momentum transfer $\mu$ in each parton-medium scattering. The effect
of the broadening on the azimuthal anisotropy can then be characterized
by the diffraction angle of each scattering $\phi_0=\mu/p_T$.
For $dE/dx=0.1(L/{\rm fm})$ GeV/fm and $\lambda\sim 1$ fm, which 
we have used to demonstrate the effect of parton energy loss in 
this paper, $\mu\sim 0.3$ GeV/$c$ according to Eq.~(\ref{eq:dedx}) 
and $\phi_0\sim 0.06$ for $p_T=5$ GeV/$c$. 
In a medium with a size of only a few mean free paths, the
contribution to $v_2$ due to such a small momentum transfer is almost
negligible. Therefore, the dominant cause of elliptic anisotropy at 
large $p_T$ will be the radiative parton energy loss in the medium.
If any suppression of large $p_T$ hadron is observed in experiments,
there should be non-vanishing azimuthal anisotropy, and vice versa.

We have considered a parton energy loss that has only a 
weak (logarithmic) dependence on the initial parton energy. 
Recent studies \cite{glv}, however, show a stronger energy dependence
as a result of the kinetic limits in the estimate of energy loss for a 
parton with finite initial energy. Such a strong energy dependence
would give a different $p_T$ dependence of hadron spectra suppression
and the accompanying azimuthal anisotropy.

For a qualitative study in this paper, we did not consider the 
expansion of the system. The longitudinal expansion will 
change the particle density in the medium, while the transverse expansion 
will decrease the spatial anisotropy as the system evolves with time. 
The decrease of particle density will reduce the average parton energy
loss over the course of the evolution. The change of spatial anisotropy
will reduce the resultant anisotropy in hadron spectra. It could
also affect the centrality dependence of the jet quenching effect. 
Detailed study of the jet quenching in a dynamically evolving system
will be the subject of future studies and is needed for a more 
quantitative analysis of any experimental data.

\acknowledgments
The author would like to thank J.-Y. Ollitrualt, A.~M.~Poskanzer 
and R.~J.~M.~Snellings
for helpful discussions. This work was supported by  
the Director, Office of Energy 
Research, Office of High Energy and Nuclear Physics, 
Division of Nuclear Physics, and by the Office of Basic Energy Science, 
Division of Nuclear Science, of  the U.S. Department of Energy 
under Contract No. DE-AC03-76SF00098 and in part by 
NSFC under project 19928511.

\end{multicols} 
\end{document}